\documentclass[aps,prb,twocolumn,superscriptaddress,floatfix,longbibliography]{revtex4-2}

\usepackage{graphicx}
\usepackage{amsmath,amssymb}
\usepackage{bm}
\usepackage{xcolor}
\usepackage[colorlinks=true,allcolors=blue]{hyperref}

\begin{document}

\title{Collective depinning of dipolar skyrmion chains:\\
thermal and athermal rounding of an elastic polymer in a quenched landscape}

\author{R. L. Silva}
\email{ricardo.l.silva@ufes.br}
\affiliation{Departamento de Ci\^{e}ncias Naturais, Universidade Federal do Esp\'{i}rito Santo, S\~{a}o Mateus, ES 29932-540, Brazil}

\author{R. C. Silva}
\email{rodrigo.c.silva@ufes.br}
\affiliation{Departamento de Ci\^{e}ncias Naturais, Universidade Federal do Esp\'{i}rito Santo, S\~{a}o Mateus, ES 29932-540, Brazil}

\author{R. L. Stamps}
\email{Robert.Stamps@umanitoba.ca}
\affiliation{Department of Physics and Astronomy, University of Manitoba, Winnipeg, Manitoba R3T 2N2, Canada}

\date{\today}

\begin{abstract}

The dynamics of a chain composed of interacting dipole skyrmions in a weak pinning potential is studied using numerical solutions of the stochastic Thiele equation. Different chain lengths with periodic boundary conditions are embedded in a quenched random landscape of attractive pinning sites and chain dynamics are examined for different applied magnetic field strengths. A well-defined threshold driving spin current is found that plays the role of the critical driving force in the context of general depinning theory. Three central results are found: (i)~The depinning threshold is independent of chain length for chains of $50$ skyrmions and longer, and the depinning field is close to one third of the value for an isolated skyrmion regardless of magnetic field. (ii)~The rounding of the transition is only about half thermal. Simulations at $T=0$ retain a width of $w(T=0)=0.334\pm0.004$ decades of drive, compared with $0.633\pm0.002$ at $300$~K, and the chain carries an intrinsic Larkin correlation length of $2.6$ bonds that temperature leaves unaltered below $k_BT\approx V_0/25$. (iii)~The width is essentially unchanged up to $k_BT\approx0.14\,V_0$, where $V_0$ is the strength of the pinning potential, and grows nearly fourfold by $k_BT\approx0.23\,V_0$, with the rounding concentrated on the creep foot rather than on the threshold itself. The width is independent of chain length at both temperatures, so the residual rounding is not a finite-size effect and no extrapolation in chain length is required to reach it. The width does depend on the applied field, but not as a simple scale --- the transition changes shape, its core widening by $26\%$ while its creep tails shorten, as the field is raised from $30$ to $40$~mT.
At chain lengths of approximately $200$, the transition proceeds through a well-defined sequence of internal deformation: growth of the Larkin length, growth of the interface roughness, and transient bond stretching, followed by recovery of the mobility and of the Hall angle. The collective reduction of the depinning threshold has direct implications for skyrmion-based devices, where chains rather than isolated textures may serve as information carriers.
\end{abstract}

\maketitle

\section{Introduction}

Magnetic skyrmions are localized, topologically nontrivial spin textures whose particle-like behavior makes them attractive candidates for information storage and logic~\cite{nagaosa2013,fert2017,woo2016,moreau2016}. To date, most studies of skyrmion depinning have focused on isolated skyrmions or dense lattices~\cite{reichhardt2015,reichhardt2022rev,lin2013}. An intermediate and physically important case, the one-dimensional chain, has received comparatively little attention, even though chains arise naturally in confined geometries and at the edges of skyrmion-hosting stripes. 

The object of the present study is the dipolar skyrmion chain, a non-worm-like-chain (non-WLC) magnetic skyrmion polymer. The mechanism binding skyrmions into a chain is dipolar interaction. Dipolar skyrmions are stabilized by the competition between the perpendicular uniaxial anisotropy, which favors uniform out-of-plane spin orientation, and the magnetostatic (shape) energy, which favors flux closure. This means that the energy of pair-wise interactions between skyrmions can be minimized by alternating helicity between neighboring skyrmions. This is possible in the dipolar chain because the helicity of a dipolar skyrmion is not fixed by a microscopic interaction so that dipolar skyrmions are free to assume either right- or left-handed textures.

A chain is qualitatively different from both an isolated skyrmion and a lattice: it is a deformable elastic object with a finite number of internal degrees of freedom, and a collective response to disorder is expected, in analogy with chains studied extensively in soft-matter systems such as polymers. Disorder pins the texture, and a finite driving force in the form of an electric current, a field gradient or thermal bias is required before motion sets in~\cite{reichhardt2017rev,iwasaki2013}. A fundamental measurable quantity for exploring this dynamic is the threshold drive for the onset of motion. A critical depinning drive can be defined for this transition and its magnitude depends on characteristic length scales of the pinning landscape and of the chain deformations~\cite{larkin1979,fisher1985,narayan1993,kolton2009}.

In our model, a spin current is used as the depinning drive. The current drives spin precession within the skyrmion chain and results in motion of the chain. This is different from field-driven dynamics because the work in field-driven dynamics is measured by the change in area enclosed by a chain, whereas for current driven skyrmion forces, work is done per chain length moved rather than area traversed. In what follows, we show how this difference affects what one expects for the magnitude of the critical driving force and explore depinning dynamics using large-scale stochastic Thiele simulations. 

We find that the depinning threshold of the chain is essentially independent of chain length at each field, with a drive value that is approximately one third of that for an isolated skyrmion. When the number of skyrmions in the chain is $50$ or more we observe collective elastic behavior where depinning is a bulk property of the chain, as is the case for the depinning of polymers. 

\begin{figure*}[tb]
\includegraphics[width=\textwidth]{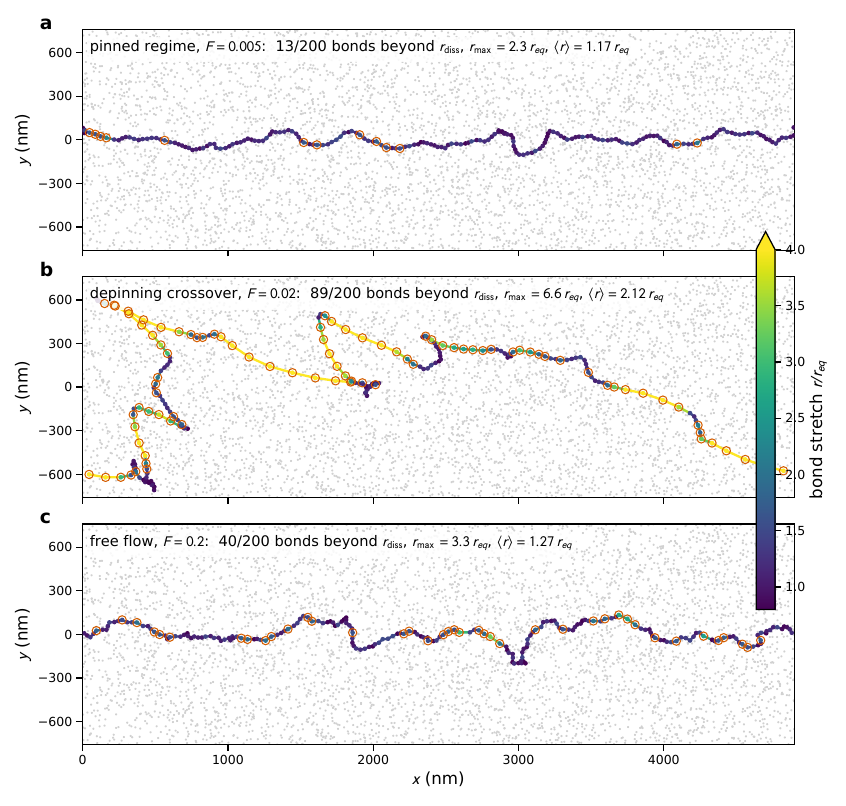}
\caption{Simulation snapshots of the $N=200$ chain at $B=35$~mT in the quenched landscape of attractive Gaussian pins (gray disks of radius $\sigma$; only a transverse window around the chain is shown), at three representative drives: (a)~pinned regime, $F=0.005$; (b)~depinning crossover, $F=0.02$; (c)~free flow, $F=0.2$. Skyrmion positions are colored by the local bond stretch $r/r_{eq}$. In the crossover the chain is rough and a large fraction of bonds is transiently stretched beyond the maximum restoring force, $r_{\rm diss}$ (yellow); in free flow the compact conformation is recovered.}
\label{fig:concept}
\end{figure*}

The transition is not sharp but rounded, and the width of that rounding is independent of chain length over the whole range studied, so it is not a finite-size effect. Thermal fluctuations account for about half of it and an athermal mechanism for the rest, as the $T=0$ simulations of Sec.~\ref{sec:T0} show directly. The depinning is therefore a rounded crossover~\cite{bustingorry2008,chauve2000} rather than a sharp critical transition, with both mechanisms contributing at room temperature; at $T=0$ the crossover survives with half the width, set by the disorder alone. Figure~\ref{fig:concept} shows, for the longest chain studied, the three regimes through which the chain passes as the drive is raised.

\section{Model and simulation methods}\label{sec:model}

The film thickness is assumed to be very small so that the dipole skyrmions are treated as two dimensional objects. This has consequences on details of how the skyrmions interact with defects, but for the dipolar skyrmions considered here, these are expected to be significant only when the film thickness approaches the size $\lambda_w$ of a domain wall. Here we assume that the thickness is less than $\lambda_w$ and that the pinning potential is sufficiently weak that the skyrmion structure remains rigid even when responding to a pinning site. This condition is satisfied by, for example, skyrmions in multilayers of  [Co/Ni]$_5$ as discussed elsewhere through atomistic calculations~\cite{silva2026chains}.

Each skyrmion is described by its center position $\bm{R}_i(t)$ and the motion of the center as determined by the Thiele equation~\cite{thiele1973,everschor2011,lin2013}
\begin{equation}
\bm{G}\times\dot{\bm{R}}_i + \alpha_D\,\dot{\bm{R}}_i = \bm{F}_i^{\rm ext}.
\label{eq:thiele}
\end{equation}
\noindent Here $\bm{G}=G_z\hat{\bm{z}}$ is the gyrovector and $\alpha_D$ the damping-dissipation coefficient (scalar). In reduced units ($a_0=1$~nm) we have $G_z=4\pi$ in all three fields, while $\alpha_D$ takes the values $1.70$, $1.10$, $0.72$ in $B=30$, $35$, $40$~mT respectively (Table~\ref{tab:params}). The three applied magnetic fields, $B=30$, $35$, and $40$~mT, are chosen so as to span a reasonably large variation of the underlying gyrotropic and damping parameters, as given in Table~\ref{tab:params}. For an isolated unpinned skyrmion under driving force $\bm{F}=F\hat{\bm{x}}$, Eq.~\eqref{eq:thiele} gives the free-flow mobility $\mu_{\rm free}$ and corresponding Hall angle $\theta_H^{\rm free}$,
\begin{equation}
\mu_{\rm free}=\frac{1}{\sqrt{\alpha_D^2+G_z^2}},\qquad
\theta_H^{\rm free}=\arctan\!\left(-\frac{G_z}{\alpha_D}\right),
\label{eq:freeflow}
\end{equation}
which serve as comparison values throughout the paper. The deterministic force is the sum of a pair interaction, a bending stiffness, the pinning force, and the drive,
\begin{equation}
\bm{F}_i^{\rm ext}=\sum_{j\neq i}\bm{F}_{ij}^{\rm pair}+\bm{F}_i^{\rm bend}+\bm{F}_i^{\rm pin}+F\hat{\bm{x}},
\end{equation}
and thermal fluctuations enter through a Langevin noise term obeying the fluctuation-dissipation relation at reduced temperature $k_BT=0.207$, which corresponds to room temperature for the [Co/Ni]$_5$ example studied here. 

\begin{table}[tb]
\caption{Model parameters at the three applied fields, in reduced units ($a_0=1$~nm). $\mu_{\rm free}$ and $\theta_H^{\rm free}$ follow from Eq.~\eqref{eq:freeflow}.}
\label{tab:params}
\begin{ruledtabular}
\begin{tabular}{lccc}
 & $B=30$~mT & $B=35$~mT & $B=40$~mT \\
\hline
$r_{eq}$ (nm) & 29.1 & 24.5 & 19.7 \\
$D_e$ & 3.10 & 3.07 & 2.94 \\
$B_{\rm rep}$ & 74.20 & 63.36 & 42.75 \\
$C_{\rm att}$ & 48.63 & 43.72 & 33.07 \\
$\delta$ (nm) & 11.68 & 10.54 & 9.89 \\
$\lambda$ (nm) & 17.53 & 15.80 & 14.84 \\
$\kappa_{\rm WLC}$ & 1.55 & 1.19 & 0.87 \\
$\alpha_D$ & 1.70 & 1.10 & 0.72 \\
$\mu_{\rm free}$ & 0.0789 & 0.0793 & 0.0794 \\
$\theta_H^{\rm free}$ & $-82.3^\circ$ & $-85.0^\circ$ & $-86.7^\circ$ \\
\end{tabular}
\end{ruledtabular}
\end{table}

The skyrmion-skyrmion interaction combines a short-range domain-wall-overlap repulsion with a longer-range magnetostatic attraction. This is modeled with the bi-exponential potential of Eq.~\eqref{eq:biexp},
\begin{equation}
U^{\rm pair}(r)=B_{\rm rep}\,e^{-r/\delta}-C_{\rm att}\,e^{-r/\lambda},\qquad \lambda>\delta,
\label{eq:biexp}
\end{equation}
whose four parameters are obtained, at each field, from atomistic Landau-Lifshitz-Gilbert relaxation of two skyrmions of opposite helicity in [Co/Ni]$_5$ multilayers~\cite{rohart2013,bessarab2015,silva2026chains}. The resulting equilibrium spacing $r_{eq}$, the binding depth $D_e$, and the bending stiffness $\kappa_{\rm WLC}$ are listed in Table~\ref{tab:params}. The ratio $\lambda/\delta\approx 1.5$ is not significantly affected by the magnetic field strengths considered here. Details concerning the choice of parameters used in the simulations presented here are provided elsewhere~\cite{silva2026chains}.

Periodic boundary conditions are used so that the depinning threshold can be identified without edge effects. This is implemented by connecting skyrmion $i$ to skyrmions $(i\pm 1)\bmod N$, thereby approximating an infinite chain. The finite length of the chain determines the size of the simulation box in the drive direction, $L_x=N\,r_{eq}$, so that all bonds are at equilibrium spacing and the closure bond does not stretch across the box. The transverse dimension of the box is fixed at $L_y=0.6\times 200\,r_{eq}$ in each field. Periodic boundary conditions are applied to the box along both axes and interactions are constrained such that each skyrmion interacts with its nearest neighbors only. The bond list is strictly topological: skyrmion $i$ interacts with skyrmions $(i\pm 1)\bmod N$ only and no distance-based neighbor search is performed, so non-bonded skyrmions that approach each other during strongly deformed transients do not interact. This simplification ensures that non-bonded skyrmion forces are not included as part of the dipolar attraction and assumes that the average density of non-bonded skyrmions is low and their effects equivalent to small, indistinguishable, perturbations to the random pinning potential. The simplification ensures that the effective nearest-neighbor parameters, listed in Table~\ref{tab:params}, remain calibrated against atomistic simulations of the full chain~\cite{silva2026chains}, in which all dipolar contributions between all skyrmions are present. In other words, long range dipolar interactions beyond the nearest neighbor can be absorbed into an effective bond energy. In the pinned and free-flow regimes the chain is locally straight and non-bonded separations are at distances $2r_{eq}$ or larger, and strongly deformed configurations remain confined to the transient crossover window.

Disorder is implemented as a frozen array of identical attractive Gaussian wells with random positions in the box, Eq.~\eqref{eq:vpin},
\begin{equation}
V^{\rm pin}(\bm{r})=-V_0\sum_{k=1}^{N_{\rm pins}}\exp\!\left[-\frac{|\bm{r}-\bm{r}_p^{(k)}|^2}{2\sigma^2}\right].
\label{eq:vpin}
\end{equation}
Each well has depth $V_0=0.5\,D_e$ and width $\sigma=0.3\,r_{eq}$, the same dimensionless ratios for all magnetic fields, giving $k_BT/V_0=0.13$--$0.14$ at each of the three fields. For the scaling analysis it is necessary to keep the pin density invariant throughout each finite-size scan. This is achieved by scaling the number of pins in proportion to the area of the simulation box such that $N_{\rm pins}=10^4\,(N/200)$, giving a fixed areal coverage $\eta_{\rm pin}\approx 11.8\%$ for every chain length and field. Each scan therefore samples statistically the same degree of disorder regardless of size.

\section{Results}
Chain transport is studied through motion of the chain's center, defined as the average position of each skyrmion in the chain. The average is done over the time series of individual skyrmion velocities in order to minimize thermal jitter. In this way, a center-of-mass drift $\bm{v}_{\rm drift}=\langle\dot{\bm{R}}_{\rm CM}\rangle_t$ is defined from the velocity time series. The drift velocity allows definition of an effective mobility $\mu_{\rm eff}=|\bm{v}_{\rm drift}|/F$ and a Hall angle $\theta_H=\arctan(\langle v_y\rangle/\langle v_x\rangle)$. Note that for small drive fields the chain will drift despite pinning due to thermal activations of skyrmions across barrier saddle points.  However this drift is one to two orders of magnitude smaller than the jitter and does not correspond to the unpinned velocity which is a monotonic function of driving field. 

Observation of the chain at different times (here referred to as snapshots) allows useful quantities to be extracted. The radius of gyration $R_g$ gives information about the degree of alignment of skyrmions; the roughness exponent $\zeta$ is a measure of how chain roughness scales with length; and the Larkin correlation length $L_c$ characterizes the elastic response of the chain to pinning. Another quantity of interest is the intact-bond fraction $f_{\rm intact}$. This is defined as the fraction of bonds with $r_i<r_{\rm diss}$ where $r_{\rm diss}$ is the distance of maximum restoring force of the bi-exponential pair potential. This maximal distance is defined by
the inflection point of the bi-exponential potential and is dependent on the strength of the magnetic field. For the parameters used in a $N=200$ chain, this quantity depends on the magnetic field as $r_{\rm diss}/r_{eq}=1.48$, $1.53$, $1.61$ at the fields $B=30$, $35$, $40$~mT.

Large-scale stochastic Thiele simulations of chains were performed. The chain lengths considered are $N=50$, $100$ and $200$, with periodic boundary conditions. Shorter chains ($N=20$) were also simulated but are not included in the analysis: their transition is markedly wider than for $N\geq50$ and they are not yet in the collective regime, see Sec.~\ref{sec:fss}. 
The scan is repeated with the same frozen potential at the three applied magnetic fields studied. We compare the chain depinning to that of an isolated skyrmion in the identical landscape at each field. 

The chain is initialized as a straight line placed uniformly along $\hat{\bm{x}}$ with spacing $r_{eq}$ at $y=L_y/2$. Equation~\eqref{eq:thiele} is then integrated for each skyrmion element of the chain with a stochastic predictor-corrector (Heun) scheme at $\Delta t=0.05$ for $1.0\times 10^8$ measurement steps after $3.3\times 10^7$ equilibration steps, sweeping the drive over $20$ logarithmically spaced values of $F$, from $3\times10^{-4}$ to $1.0$ at $300$~K and from $5\times10^{-3}$ to $0.3$ at $T=0$, where the transition sits at higher drive. At each force we run $11$ independent disorder realizations, with $16$ thermal replicas per realization at finite temperature and a single deterministic trajectory at $T=0$.
Each mobility curve therefore rests on $11$ realizations at each of $20$ drives, that is $220$ realization-force pairs; at finite temperature each pair carries $16$ replicas, so $3520$ individual trajectories. Statistical uncertainties on quantities derived from it ($F_c$, $w$, $\zeta$) are estimated by bootstrap resampling over disorder realizations, with the replica average taken within each realization. All tabulated quantities and all figures except the temperature scan of the Larkin length in Sec.~\ref{sec:T0} use this data set. The temperature scan of the Larkin length in Fig.~\ref{fig:T0}(c) follows a separate and lighter protocol, because it is read at fixed drive rather than across the transition: at each of the six intermediate temperatures we sweep five drives inside the pinned window ($F\leq 0.005$) with five disorder realizations each, that is $25$ runs per temperature. For each run $L_c$ is averaged over the second half of the recorded frames, the first quarter being equilibration from the straight initial chain; the value quoted for a temperature is the mean over the five drives and the uncertainty is the standard error between them, which exceeds the scatter within a single drive. The $T=0$ and $300$~K anchors of that panel are read from the sweep above with the same estimator.

The equilibration does not remove all memory of the straight initial condition. Splitting the measurement window into halves and repeating the threshold analysis on each, $F_c^{(50)}$ rises by $3\%$ from the first to the second half at $B=35$ and $40$~mT ($0.0349\to0.0359$ at $B=35$~mT, $N=200$) and by $5\%$ at $T=0$, while $w$ changes by $1\%$ at $B=35$ and $40$~mT and by $10\%$ both at $T=0$ ($0.34\to0.31$) and at $B=30$~mT ($0.73\to0.66$), where the creep foot is still relaxing. All values quoted below use the full measurement window; the second-half thresholds are listed in Table~\ref{tab:summary} as a check. The residual drift is of the order of the field-to-field differences in $w$ discussed in Sec.~\ref{sec:fss} and is kept in mind there. Velocities are the drift velocity $|\langle\dot{\bm R}_{\rm CM}\rangle_t|$ throughout; the modulus average $\langle|\dot{\bm R}_{\rm CM}|\rangle_t$ gives the same thresholds to within $1\%$ but a creep floor two to four times higher, because it counts thermal excursions that do not transport the chain.

For comparison with the isolated-skyrmion limit, we ran an $N=1$ benchmark in the same box as the $N=200$ chain, with the same $10^4$ pins and hence the same coverage, in each field and at both temperatures, with the same protocol and statistics ($11$ realizations, $16$ thermal replicas at $300$~K and one deterministic trajectory at $T=0$); its force grid spans $0.02$ to $0.4$, where the single-skyrmion threshold lies.

\subsection{The depinning transition region}\label{sec:overview}

A chain length of $N=200$ was used to identify depinning and pinning parameters chosen to satisfy two competing requirements: (i)~a well-developed sub-mobility plateau in the pinned regime and (ii)~a chain that remains structurally intact through the depinning transition. Deeper pins where $V_0\gtrsim D_e$, with $D_e$ being the energy gained by forming a bond between two skyrmions, produce a stronger plateau but rupture the chain at the transition. The length $N=200$ allows the configuration to deform sufficiently to a high density of weak pins so that the cumulative drag is strong and the chain remains connected. Note that bonds between skyrmions can stretch beyond $r_{\rm diss}$ during depinning. The stretching is transient: the bond list is fixed by the chain topology, so a stretched pair remains connected and recovers its equilibrium separation once the inhomogeneous drag of the landscape disappears in free flow. Permanent separation of a pair would require an energy larger than $D_e$, which is not reached for the parameters used here (as noted earlier). This transient stretching and subsequent recovery is shown explicitly in the movie provided as Supplemental Material~\cite{SM}, which follows the same $N=200$ chain at $B=35$~mT in the pinned, crossover, and free-flow regimes.

\begin{figure*}[tb]
\includegraphics[width=\textwidth]{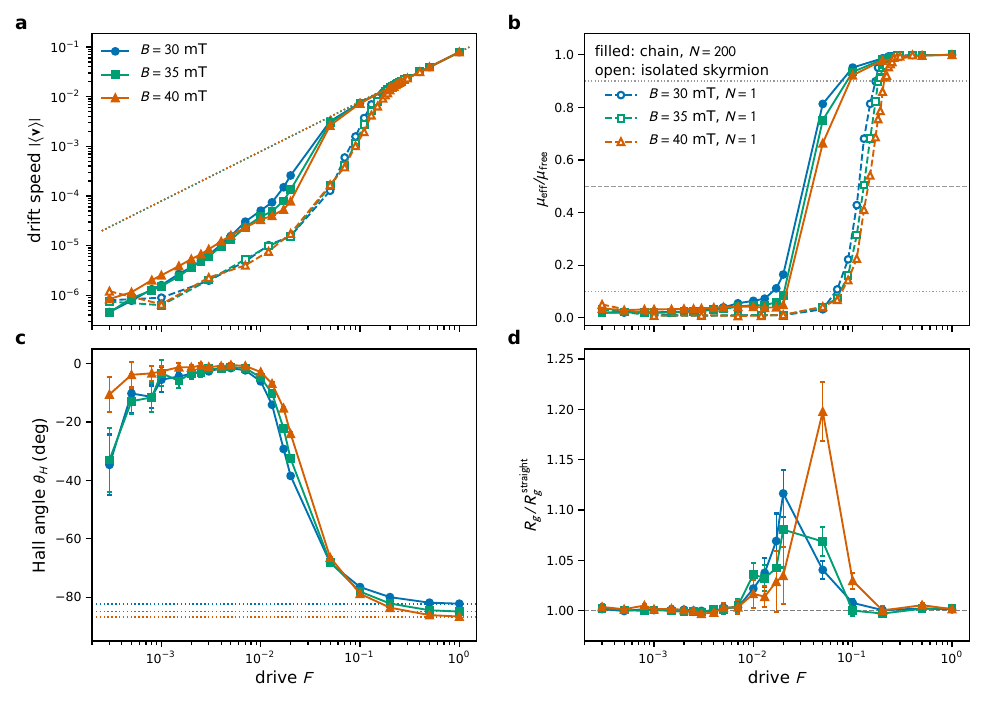}
\caption{Transport observables of the chain (solid markers, $N=200$) and the isolated skyrmion (open/dashed) versus drive at the three fields. (a)~Drift velocity in log-log scale; dotted lines: free-flow law for each field. (b)~Mobility ratio $\mu_{\rm eff}/\mu_{\rm free}$ in lin-log. (c)~Hall angle of the chain; dotted lines mark the predicted free-flow values. (d)~Radius of gyration normalized to the straight-chain reference $R_g^{\rm straight}=(N-1)r_{eq}/\sqrt{12}$, showing transient stretching at the transition. Error bars span the $16$th to $84$th percentiles of a bootstrap over disorder realizations, with the thermal-replica average taken within each realization (the same protocol as in Figs.~\ref{fig:fss}--\ref{fig:tscan}); in (a) and (b) they are smaller than the symbols in the pinned and free-flow regimes and become comparable to the symbol size inside the crossover window, where fluctuations are largest.}
\label{fig:overview}
\end{figure*}

Figure~\ref{fig:overview} shows transport parameters for three different magnetic fields as a function of $F$: (a) the drift velocity, (b) the effective mobility, (c) the Hall angle, and (d) the radius of gyration of the chain. The chain length is $N=200$; results are shown for the chain and the isolated-skyrmion benchmark for each field. 

Three regimes are visible in all fields for each parameter. At low drive ($F\lesssim 5\times 10^{-3}$) the chain and isolated skyrmion each creep with strongly reduced linear mobility. This creep region defines plateaus in (b) for the mobility ratio: $\mu_{\rm pinned}/\mu_{\rm free}\approx 0.02$--$0.04$ for the chain, and $\approx 0.01$--$0.02$ for the isolated skyrmion, whose drift at the lowest drives ($F\lesssim 10^{-3}$) is limited by the thermal noise floor of a single particle. At high drive ($F\gtrsim 0.2$) free-flow mobility is recovered to within $5\%$ of that predicted by Eq.~\eqref{eq:freeflow}. The mobility ratio shown in (b) displays the same features as the drift velocity but with a more pronounced initial threshold and thermal rounding for the final transition out of the creep regime into a flow regime. The Hall angle of the chain [Fig.~\ref{fig:overview}(c)] matches the single-skyrmion value $\theta_H^{\rm free}$ within $1^\circ$ at high drive for all magnetic fields. Note also that the thermally rounded depinning crossover exists between these limits also for the isolated skyrmion, although with a higher $F$ onset.

The Hall angle for the chain shown in Fig.~\ref{fig:overview}(c) as a function of $F$ likewise shows a plateau and thermal rounding. The Hall angle provides an independent, transport-free diagnostic of the depinning~\cite{jiang2017,litzius2017}. In the pinned regime the chain's lateral motion is locked by the landscape and $\theta_H\approx 0^\circ$. In free flow $\theta_H$ recovers the single-skyrmion value $\theta_H^{\rm free}$ predicted by Eq.~\eqref{eq:freeflow}, matching theory within $1^\circ$ at every field.  At the lowest drives there are additional features that can be interpreted as changes in shape of the chains. For comparison, the Hall angle for perfectly straight chains is shown by horizontal dashed lines for each field. The rotation of $\theta_H$ from $0^\circ$ to its asymptote occurs in the same drive window as the mobility rise. The chains are very far from flat in the lowest $F$ region with correspondingly large fluctuations around zero Hall angle and in (a) and (b) large fluctuations in drift velocities and mobility ratios.

The radius of gyration ($R_g$) is plotted as a function of $F$ in (d). $R_g$ has been normalized to the straight-chain reference $R_g^{\rm straight}=r_{eq}\sqrt{(N^2-1)/12}\approx Nr_{eq}/\sqrt{12}$, the radius of gyration of $N$ points evenly spaced along a line (the chain is unwrapped from the periodic box before $R_g$ is computed). There are large fluctuations throughout the transition region which are also strongly dependent on magnetic field strength.
With bonds held near $r_{eq}$ a chain cannot exceed the straight-rod value, so excursions of $R_g$ above unity are possible only through genuine bond stretching, which makes the transient peak at the crossover a direct signature of stretched bonds rather than mere roughening. In the pinned and free-flow regimes $R_g/R_g^{\rm straight}$ stays within $0.3\%$ of unity at all fields. The peak excursion is $12\pm2\%$, $8\pm4\%$ and $20\pm3\%$ at $B=30$, $35$ and $40$~mT (mean and standard error over disorder realizations, averaged over the second half of the run), located at $F=0.02$ for the two lower fields and at $F=0.05$ for $B=40$~mT; the large realization-to-realization scatter reflects the intermittent character of the stretching events.

\subsection{Collective chain behaviour and scaling}\label{sec:fss}

An operational depinning force $F_c(\theta)$ is defined as the largest drive at which $\mu_{\rm eff}/\mu_{\rm free}$ crosses a threshold value $\theta$ from below. Central values are read from the mobility curve averaged over all disorder realizations, and uncertainties are the $16$th and $84$th percentiles of a bootstrap over those realizations. Where the bootstrap distribution is asymmetric the two percentiles are quoted separately; this happens only when the creep valley approaches the lower threshold, and at $N\geq 50$ only for $B=30$~mT at $N=50$. We also report a reduced threshold, defined as the ratio of the chain and isolated-skyrmion thresholds, $F_c^{\rm chain}/F_c^{(1)}$. A linear interpolation of $\log F$ is used to determine the threshold in the presence of statistical noise in the creep regime. A range of $\theta$ values between $0.3$ and $0.7$ was tested. The thresholds obtained for an $N=200$ chain were statistically independent of thermal noise for all values of $\theta$ in this range.

The threshold values were also found to be independent of magnetic field. Choosing $\theta=0.5$, we measured $F_c(N=200)/F_c(N=1)=0.279\pm0.004$, $0.273\pm0.006$ and $0.273\pm0.006$ at $B=30$, $35$ and $40$~mT, where the uncertainties were estimated from the disorder realizations using bootstrap sampling. The three values coincide with one another within $2.2\%$. Note that because the force grid is logarithmic and the $\theta=0.5$ crossing falls between grid points, the central values carry a scheme-dependent shift of order $+0.03$ if $F$ rather than $\log F$ is interpolated (this shifts the ratio to $0.29$--$0.31$). The ratio is therefore $\approx 0.3$, consistent with a ratio of one third.

\begin{figure*}[tb]
\includegraphics[width=\textwidth]{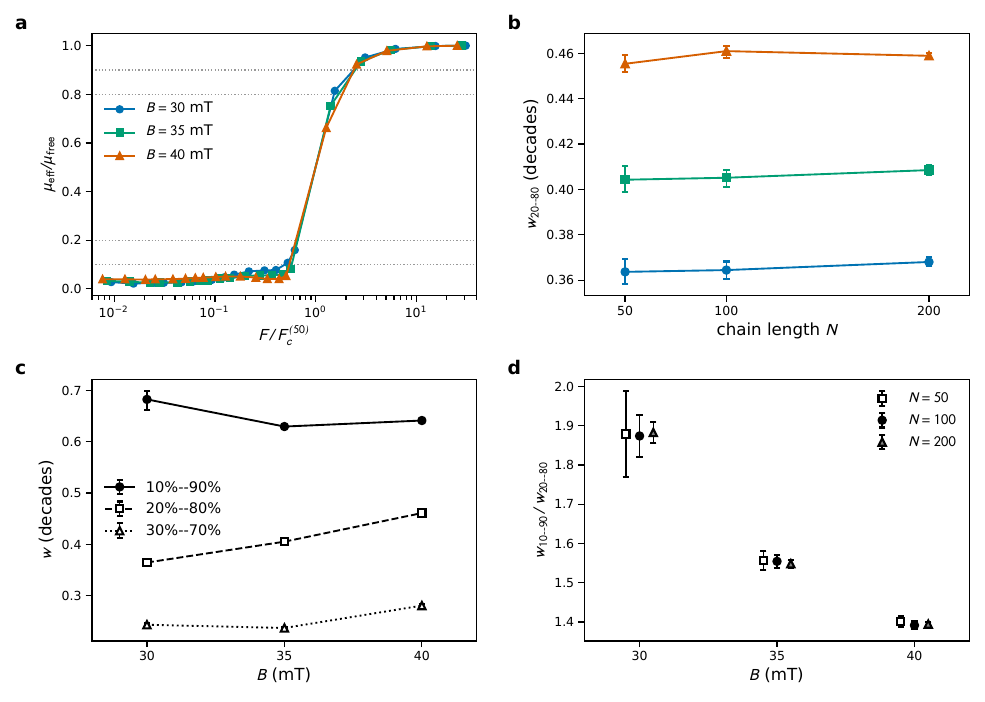}
\caption{Field dependence of the depinning transition at $300$~K. Panels (a) and (c) use $N=100$; panel (d) shows all three chain lengths. (a)~Mobility ratio versus drive, with the drive rescaled by $F_c^{(50)}$ of each field. The three curves collapse above threshold but not below it: the creep foot of $B=30$~mT sits well above that of $B=40$~mT, so the transitions differ in shape and not only in scale. Dotted lines mark the $10\%$ and $90\%$ levels, dashed lines the $20\%$ and $80\%$ levels. (b)~The $20\%$--$80\%$ width versus chain length at each field: flat in $N$. (c)~The three widths versus field. They do not rank the fields in the same order, which is only possible if the curve shape changes with field. (d)~The ratio $w_{10\text{--}90}/w_{20\text{--}80}$, measuring how much of the transition lies in the tails relative to the core: monotonic in field and independent of chain length, shown for all three sizes with a small horizontal offset so that the coincident points remain visible.}
\label{fig:fss}
\end{figure*}

To test how this ratio and the depinning threshold depend on the periodicity enforced by the periodic boundary conditions, we ran the chain at $N=50$, $100$, $200$ in each of the three fields, holding the pin density and all landscape parameters fixed. A summary of the results is presented in Fig.~\ref{fig:fss}.

\begin{table}[tb]
\squeezetable
\caption{Operational depinning forces of the chain at the three fields. $F_c(N=1)$ is the isolated-skyrmion benchmark at $\theta=0.5$. The ratio $F_c(N=200)/F_c(N=1)$ is close to $1/3$ at every field. $F_c(\theta)$ is the largest-$F$ crossing of $\mu_{\rm eff}/\mu_{\rm free}=\theta$ with log-$F$ interpolation; parenthesized digits are the half-width of the $16$th--$84$th percentile interval of a bootstrap over disorder realizations. The width $w(N)=\log_{10}[F_c(90\%)/F_c(10\%)]$ is given in decades and is independent of $N$ over the range listed. The row ``2nd half of window'' gives $F_c^{(50)}(N=200)$ evaluated on the second half of the measurement window only (Sec.~\ref{sec:model}).}
\label{tab:summary}
\begin{ruledtabular}
\begin{tabular}{lccc}
 & $B=30$~mT & $B=35$~mT & $B=40$~mT \\
\hline
$F_c(N=1)$ & 0.1153(15) & 0.1295(31) & 0.1433(36) \\
$F_c^{(50)}$, $N=50$  & 0.0323(3) & 0.0354(2) & 0.0390(4) \\
$F_c^{(50)}$, $N=100$ & 0.0322(2) & 0.0354(1) & 0.0392(3) \\
$F_c^{(50)}$, $N=200$ & 0.0321(1) & 0.0354(1) & 0.0392(2) \\
\quad 2nd half of window & 0.0329 & 0.0359 & 0.0396 \\
$w$, $N=50$  & $0.683^{+0.048}_{-0.028}$ & 0.629(3) & 0.638(3) \\
$w$, $N=100$ & $0.683^{+0.016}_{-0.021}$ & 0.630(3) & 0.642(2) \\
$w$, $N=200$ & 0.693(9) & 0.633(2) & 0.640(1) \\
\end{tabular}
\end{ruledtabular}
\end{table}

Two features stand out and are robust with respect to the applied field. First, consider the $50\%$ depinning threshold, listed in Table~\ref{tab:summary}. This threshold is denoted $F_c^{(50)}$ and is the drive at which the disorder-averaged mobility ratio reaches half of its free-flow value, $\mu_{\rm eff}/\mu_{\rm free}=0.5$. It is essentially $N$-independent in all fields, with a spread below $2\%$ between $N=50$ and $N=200$. 

Combined with $F_c(N=1)$ in each field, this gives a collective-advantage ratio

\begin{equation}
\frac{F_c^{(50)}(N\geq 50)}{F_c^{(50)}(N=1)}\approx \frac{1}{3},
\label{eq:onethird}
\end{equation}

\noindent that is a property of the collective elastic regime and is already realized at $N=50$. This ratio is robust both against the threshold $\theta$ chosen to define $F_c$ (within $\theta\in[0.3,0.7]$) and against the applied field over $B=30$--$40$~mT, even though the underlying damping $\alpha_D$ varies by more than a factor of two over this range. The reduction of the depinning threshold by a factor of three reflects the chain's ability to redistribute forces across the chain. Skyrmions experiencing pinning forces can be pulled away from the pinning site by skyrmions between pinning sites that are able to move more freely in response to $F$. In this way the chain advances as a whole at driving $F$ that would not move an isolated pinned skyrmion.

Second, consider the width of the transition, defined by the separation in drive between the crossings of two chosen mobility levels. We take the levels $10\%$ and $90\%$ and define the width as $w(N)=\log_{10}[F_c(90\%)/F_c(10\%)]$, the number of decades of drive between the $10\%$ and $90\%$ mobility marks. Where the estimator is well defined, the width does not depend on the chain length. At $B=35$~mT, $w=0.629\pm0.003$, $0.630\pm0.003$ and $0.633\pm0.002$ at $N=50$, $100$ and $200$, each from the full set of $220$ realization-force pairs, constant with $\chi^2/{\rm dof}=0.6$; a fit $w(N)=w_\infty+a/N$ returns a slope consistent with zero, and no extrapolation in $N$ is required to reach the long-chain value.

The width is defined only where the mobility curve crosses the $10\%$ mark from below, and at finite temperature it need not do so. Thermal creep gives the chain a finite velocity at arbitrarily small drive, so the ratio $\mu_{\rm eff}/\mu_{\rm free}=v/(F\mu_{\rm free})$ diverges as $F\to 0$: the curve descends from that divergence into a creep valley and only then rises through the transition. The $10\%$ level is read on the near edge of that valley, and the reading is meaningful only while the valley floor lies clearly below it. The floor rises as the chain is shortened, a shorter chain being carried over the landscape by thermal activation more easily: at $B=35$~mT it is $0.019$ for $N=200$ and $0.027$ for $N=20$ (drift velocity). The $N=20$ chain is excluded from the width analysis for a different reason: its width, $w=0.84$ decades at $B=35$~mT against $0.63$ for $N\geq50$, shows that it is not yet in the collective regime, consistent with the Larkin length of $\approx4$ bonds no longer being small compared with the chain. For $N\geq 50$ every bootstrap resample yields a crossing at all three fields.
 For $N\geq50$ the floor lies at $0.02$--$0.04$ at all three fields, well below the $10\%$ level, but the crossing is least well determined at $B=30$~mT, where the mobility curve leaves the creep valley most gradually and is still relaxing within the measurement window (Sec.~\ref{sec:model}). The consequence is visible in the error bars of Table~\ref{tab:summary}: at $B=30$~mT the width at $N=50$ carries an uncertainty of $^{+0.05}_{-0.03}$ decades, against $\pm0.003$ at $B=35$~mT for the same statistics. This is the same failure mode that appears at fixed $N$ when the temperature is raised instead, and which is documented in Sec.~\ref{sec:tscan} at $T=510$~K, where the foot falls below the accessible drive window altogether. That the exclusion is a property of the estimator and not of the physics can be checked by moving the thresholds clear of the valley: measured between $20\%$ and $80\%$, or between $30\%$ and $70\%$, the width is independent of chain length down to $N=20$ at all three fields.

Across fields the width is not constant, and the way it varies depends on where the transition is read. At $N=100$, with the full set of $220$ realization-force pairs behind every curve, the $20\%$--$80\%$ width is $0.364\pm0.004$, $0.405\pm0.004$ and $0.461\pm0.003$ decades at $B=30$, $35$ and $40$~mT: a monotonic increase of $26\%$, far outside the uncertainties. The $10\%$--$90\%$ width, by contrast, gives $0.683^{+0.016}_{-0.021}$, $0.630\pm0.003$ and $0.642\pm0.002$, which does not order the fields at all, and the $30\%$--$70\%$ width gives $0.243\pm0.003$, $0.237\pm0.002$ and $0.280\pm0.004$, ordering them differently again.

Three definitions of the same quantity cannot rank the same three curves in different orders unless the curves differ in shape and not merely in scale. Figure~\ref{fig:fss}(a) shows this directly: rescaling the drive by $F_c^{(50)}$ of each field collapses the three mobility curves above threshold but not below it, where the creep foot of $B=30$~mT lies well above that of $B=40$~mT. A single number captures the difference. The ratio $w_{10\text{--}90}/w_{20\text{--}80}$, which measures how much of the transition sits in the tails relative to the core, falls monotonically from $1.87^{+0.05}_{-0.06}$ at $B=30$~mT to $1.55\pm0.02$ at $B=35$~mT and $1.39\pm0.01$ at $B=40$~mT, see Fig.~\ref{fig:fss}(d). It does not depend on chain length: at $N=50$ the same three fields give $1.88^{+0.14}_{-0.08}$, $1.56\pm0.02$ and $1.40\pm0.01$.

The effect is not a thermal one in disguise. The reduced temperature $k_BT/V_0$ changes by only $5\%$ across the three fields, from $0.134$ to $0.141$, while the core width changes by $26\%$ and the shape ratio by $36\%$. What does change substantially is the dynamics: over the same range the damping $\alpha_D$ falls by a factor of $2.4$ and the bending stiffness $\kappa_{\rm WLC}$ by $1.8$. A less damped, more flexible chain depins through a transition with a wider core and shorter creep tails.

The foot of the transition is likewise stable once the estimator is well posed. Over $N=50$ to $200$ the low-end threshold $F_c(10\%)$ varies by at most a factor of $1.3$, and at $B=35$~mT, where each point rests on $220$ runs, it is constant to $1\%$: $F_c(10\%)=0.0205$, $0.0205$ and $0.0204$ at $N=50$, $100$ and $200$. The strong apparent drift of the foot with $N$ reported for shorter chains is an artifact of reading the $10\%$ level inside the creep valley.

\subsection{Internal structure of the chain}\label{sec:micro}

Figure~\ref{fig:concept} shows the chain in the three regimes directly: locally straight and pinned at $F=0.005$, rough and transiently stretched inside the crossover at $F=0.02$, and compact again in free flow at $F=0.2$. As noted earlier, the chain outside the pinned creep regime is straight. In the pinned regime, the chain remains essentially straight with $R_g$ equal to its straight-chain value as seen in Fig.~\ref{fig:overview}(d). At the crossover, $R_g$ peaks transiently above the straight-line reference, signaling internal stretching driven by the inhomogeneous drag of the landscape with sections that have cleared a pin advancing while sections that remain trapped lag behind, so that the bonds connecting them stretch beyond $r_{eq}$. The peak excursion is largest at $B=40$~mT ($20\%$), where the binding is weakest, but does not order monotonically with field ($12\%$ and $8\%$ at $30$ and $35$~mT), see Sec.~\ref{sec:overview}.

\begin{figure*}[tb]
\includegraphics[width=\textwidth]{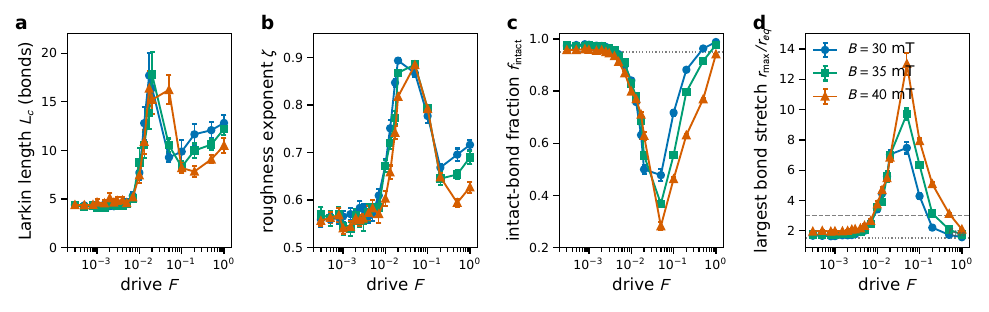}
\caption{Internal structure of the chain at the three fields ($N=200$). (a)~Larkin correlation length $L_c$ versus drive: an intrinsic $\approx 4$ bonds in the pinned regime, a sharp peak at $F\approx 0.02$, and relaxation in free flow. (b)~Roughness exponent $\zeta$, peaking in the same $F$ window. (c)~Intact-bond fraction $f_{\rm intact}$, dipping deeply in the same window (minima $0.48$, $0.37$, $0.28$ at $B=30$, $35$, $40$~mT) and recovering toward unity in free flow; the dotted line marks $f_{\rm intact}=0.95$. (d)~Largest bond stretch $r_{\max}/r_{eq}$ in the chain; dotted line $r_{\rm diss}$, dashed line $3\,r_{eq}$. All quantities are averaged over the second half of the recorded configurations, after the chain has left its straight initial condition, and over eleven disorder realizations (error bars: standard error). All four signal that the chain deforms internally before recovering macroscopic mobility.}
\label{fig:micro}
\end{figure*}

The microscopic content of this stretching is captured by the Larkin correlation length $L_c$ and the roughness exponent $\zeta$. Results for these parameters are shown in Fig.~\ref{fig:micro}(a) and (b), respectively. In the pinned regime $L_c$ sits at an intrinsic value of $\approx 4$ bonds at every field.
At the crossover this value rises sharply to a peak at $F\approx 0.02$, reaching $17.7\pm2.4$, $17.8\pm2.3$ and $16.4\pm1.8$ bonds at $B=30$, $35$ and $40$~mT (standard error over disorder realizations), and then relaxes in the flow regime to $10$--$12$ bonds, well above the pinned-regime value: in free flow the chain is smooth on the scale of the pin spacing, so transverse correlations extend further than in the pinned state. The roughness exponent behaves similarly, rising from $\zeta\approx 0.56$--$0.57$ in the pinned regime to a peak $\zeta\approx 0.88$--$0.89$ at $F\approx 0.02$--$0.05$, and settling at $0.61$--$0.71$ in free flow.

The intact-bond fraction $f_{\rm intact}$ [Fig.~\ref{fig:micro}(c)] quantifies how the chain restructures. In the pinned regime thermal stretching keeps $f_{\rm intact}$ at approximately $0.95$--$0.97$. On approach to the transition $f_{\rm intact}$ falls below $0.95$ at $F\approx 0.005$--$0.007$, which is before the mobility midpoint, and reaches a deep minimum at $F=0.05$, inside the crossover window, where $f_{\rm intact}^{\rm min}=0.48\pm0.02$, $0.37\pm0.01$ and $0.28\pm0.02$ at $B=30$, $35$, and $40$~mT. The largest bond in the chain, averaged over realizations, reaches $7.5$, $9.7$ and $13\,r_{eq}$ at the same drive [Fig.~\ref{fig:micro}(d)], with individual configurations reaching $11$, $15$ and $23\,r_{eq}$; at these separations the bi-exponential attraction is negligible, so the pair is held together only by the fixed bond topology of the model, and it is the loss of the differential drag in free flow, not the bond force, that brings the two sections back together. As discussed earlier, this means that a large fraction of the bonds are stretched beyond the maximum restoring force distance during depinning but the pinning parameters used in the simulations are such that motion across a pinning site is still possible due to interactions between skyrmions. Thus $f_{\rm intact}$ recovers toward unity in the free flow regime, $f_{\rm intact}=0.99$, $0.98$ and $0.94$ at $F=1$ for $B=30$, $35$ and $40$~mT, with the largest bond still at $1.7$--$2.6\,r_{eq}$. The depinning of the chain is not a rigid-object process but proceeds through massive bond deformation which is most pronounced at higher fields where the binding is weakest.

The movie included as Supplemental Material~\cite{SM} makes this sequence directly visible. It follows a single disorder realization of the $N=200$ chain at $B=35$~mT through the three regimes, at $F=0.005$, $0.05$ (the minimum of $f_{\rm intact}$) and $1$, i.e.\ deeper into the crossover and into free flow than the snapshots of Fig.~\ref{fig:concept}: in the pinned regime the chain is locally straight and only a handful of bonds ($7$ of $200$ in the frames shown) exceed $r_{\rm diss}$; inside the crossover window the chain develops large transverse excursions and more than half of the bonds ($114$ of $200$) are simultaneously stretched beyond $r_{\rm diss}$; in free flow the compact, locally straight conformation is recovered and the count returns to a few bonds ($4$ of $200$). The chain retains its connectivity throughout --- the bond list is topological and fixed --- so that the deep excursion of $f_{\rm intact}$ registers transient, reversible stretching rather than permanent fragmentation.

All three structural observables peak (or dip) in the same $F$ window and ahead of the midpoint of the mobility rise and of the Hall-angle recovery. The chain deforms first and recovers once it is finally depinned. The intrinsic $L_c$ and the peak $\zeta$ are field-independent, and the peak $L_c$ is the same at the three fields within its uncertainty ($\approx2$ bonds).

\subsection{Temperature dependence}\label{sec:tscan}

\begin{figure}[tb]
\includegraphics[width=\columnwidth]{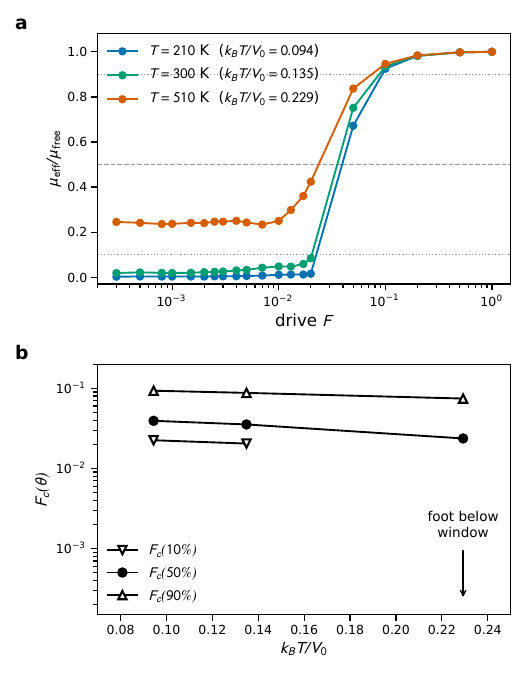}
\caption{Temperature dependence of the depinning transition at fixed landscape ($B=35$~mT, chain $N=200$). (a)~Mobility ratio versus drive at $T=210$, $300$ and $510$~K ($k_BT/V_0=0.094$, $0.135$, $0.229$); dotted and dashed lines mark the $10\%$, $50\%$ and $90\%$ levels. As $T$ increases the creep foot lifts and the transition shifts to lower drive. (b)~Operational thresholds $F_c(10\%)$, $F_c(50\%)$ and $F_c(90\%)$ versus $k_BT/V_0$. The foot $F_c(10\%)$ drops fastest and falls below the accessible drive window at $T=510$~K (arrow), while the center and the upper end shift only mildly; the widening separation between foot and shoulder is the thermal growth of the residual width $w$.}
\label{fig:tscan}
\end{figure}

The width that survives at every chain length must have a source other than the chain length. We test the thermal part of it directly by repeating the scan at fixed landscape ($B=35$~mT, chain length $N=200$) at two further temperatures, $T=210$ and $510$~K, bracketing the reference $T=300$~K; in reduced units these correspond to $k_BT/V_0=0.094$, $0.135$ and $0.229$. Figure~\ref{fig:tscan}(a) shows the mobility curves at $N=200$.

Two effects are visible. First, the $50\%$ threshold decreases monotonically with temperature, with $F_c^{(50)}=0.0393$, $0.0354$ and $0.0237$ at $T=210$, $300$ and $510$~K, as thermal activation assists the chain over the pins. 

The second point, which is central to the interpretation of the residual width, is seen from Fig.~\ref{fig:tscan}(b).  Between $210$ and $300$~K the transition shifts almost rigidly: the foot drops from $F_c(10\%)=0.0225$ to $0.0204$ and the center from $F_c(50\%)=0.0393$ to $0.0354$, by $9\%$ and $10\%$ respectively, while $F_c(90\%)$ changes by less than $8\%$. The rounding from below becomes visible only at the highest temperature: at $T=510$~K the creep valley no longer descends below $\mu_{\rm eff}/\mu_{\rm free}\approx 0.24$, so the $10\%$ mark is never crossed and the foot lies below our force window altogether, while the center has moved by only a further factor of $1.5$.

The width therefore grows with $k_BT/V_0$, as expected for a thermally rounded transition, but not uniformly: $w=0.62$ and $0.63$ decades at $T=210$ and $300$~K, a difference at the level of our uncertainties, and $w\gtrsim 2.4$ decades at $T=510$~K, the last value a lower bound set by the foot falling below the accessible drive. The width is thus essentially unchanged over $k_BT/V_0=0.094$--$0.135$ and grows by a factor of at least $3.8$ by $0.229$ ($w\gtrsim2.4$ against $0.63$). At that temperature the rounding is asymmetric, in that temperature affects the creep foot of the transition strongly while leaving the bulk threshold $F_c^{(50)}$ and the upper end only weakly shifted. The foot of the transition is therefore largely defined by thermally assisted depinning. How much of the residual width this accounts for is settled in Sec.~\ref{sec:T0}, by setting the temperature to zero.

\subsection{The athermal limit}\label{sec:T0}

The temperature scan above establishes that the residual width grows with $k_BT/V_0$, but it cannot say how much of the width at room temperature is thermal, because even the coldest temperature reached still rounds the transition. We therefore repeated the size scan at $T=0$, with the Langevin term switched off and every other element of the protocol unchanged: same field $B=35$~mT, same landscape parameters, same force grid, same estimator.

\begin{figure*}[tb]
\includegraphics[width=\textwidth]{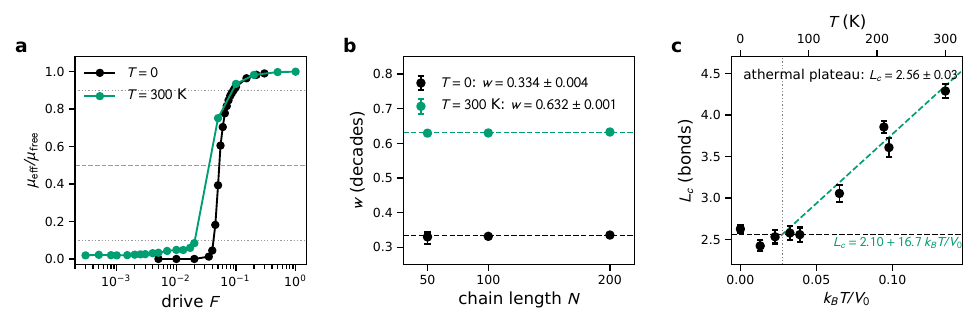}
\caption{The athermal limit, at $B=35$~mT. (a)~Mobility ratio versus drive for the $N=200$ chain at $T=0$ and $300$~K. At $T=0$ the curve rises from exactly zero and is monotonic; at $300$~K thermal creep gives a finite velocity at arbitrarily small drive, so $\mu_{\rm eff}/\mu_{\rm free}=v/(F\mu_{\rm free})$ diverges as $F\to 0$ and the curve descends into a creep valley before rising through the transition. Dotted lines mark the $10\%$ and $90\%$ levels; the $10\%$ level is read on the near edge of the valley, which is what restricts the width analysis to $N\geq 50$. (b)~Transition width versus chain length at the two temperatures: both constant in $N$ (dashed lines: weighted means, $0.632\pm0.001$ and $0.334\pm0.004$), with no drift to extrapolate away. The gap between the bands is the thermal contribution, $47\%$ of the room-temperature width. (c)~Larkin correlation length in the pinned regime ($F\le0.005$) versus temperature, from the transverse displacement correlation of the $N=200$ chain. $L_c$ holds at $2.56\pm0.03$ bonds from $T=0$ to $87$~K ($k_BT/V_0=0.039$) and grows linearly from $145$~K upward as $L_c=2.10+16.7\,k_BT/V_0$; the two lines meet at $k_BT/V_0\approx0.03$ (dotted vertical line). Upper axis: temperature in kelvin for the $[$Co/Ni$]_5$ parameters.}
\label{fig:T0}
\end{figure*}

At $T=0$ the mobility curve is qualitatively different [see Fig.~\ref{fig:T0}(a)]. Below threshold the chain does not move at all, so $\mu_{\rm eff}/\mu_{\rm free}$ rises from exactly zero and the curve is monotonic in $F$. The creep valley that complicates the $10\%$ crossing at $300$~K has no counterpart here, and the width is well defined at every chain length.

The width is again independent of $N$. We measure $w=0.330\pm0.016$, $0.331\pm0.007$ and $0.335\pm0.005$ decades at $N=50$, $100$ and $200$, each from the full set of $220$ realization-force pairs, consistent with a constant
\begin{equation}
w(T=0)=0.334\pm0.004~{\rm decades},
\label{eq:wT0}
\end{equation}
with $\chi^2/{\rm dof}=0.11$; the coefficient of a $1/N$ term is $-0.4\pm1.0$, consistent with zero. Against $w(300~{\rm K})=0.633\pm0.002$ at the same field, Fig.~\ref{fig:T0}(b), the ratio is $1.90\pm0.02$. Finite temperature therefore accounts for $47\%$ of the rounding at room temperature, and an athermal mechanism accounts for the remainder.

Whether that width belongs to the ensemble or to a single sample can be settled directly, because at $T=0$ each disorder realization is deterministic. Measuring $w$ from the mobility curve of each realization on its own, and comparing with the width of the ensemble-averaged curve, gives Table~\ref{tab:T0single}.

\begin{table}[tb]
\caption{Athermal ($T=0$) transition width at $B=35$~mT, measured within a single disorder realization and from the ensemble-averaged mobility curve. The quoted spread is over the eleven realizations.}
\label{tab:T0single}
\begin{ruledtabular}
\begin{tabular}{ccc}
 $N$ & $w$, single realization & $w$, ensemble average \\
\hline
  50 & $0.316\pm0.044$ & $0.331$ \\
 100 & $0.325\pm0.030$ & $0.333$ \\
 200 & $0.332\pm0.023$ & $0.336$ \\
\end{tabular}
\end{ruledtabular}
\end{table}

The two columns of Table~\ref{tab:T0single} agree. The width does not arise from averaging over landscapes: it is already present in one of them, and it would not narrow if the number of realizations were increased without bound.

The part that does come from the ensemble is small. Treating the two contributions as adding in quadrature, the ensemble part is $0.10$, $0.07$ and $0.05$ decades at $N=50$, $100$ and $200$, while the intrinsic part is unchanged at $0.32$--$0.33$. Each of these ensemble values is compatible with zero once the spread over realizations is propagated, so we quote them as upper bounds, consistent with but not by themselves establishing the $N^{-1/2}$ decay expected from self-averaging. Self-averaging is seen directly, and with significance, in the position of the transition: the scatter $\sigma(F_c^{(50)})/F_c^{(50)}$ falls from $4.8\%$ to $2.8\%$ over the same range. The chain does self-average; what self-averages is where the transition sits, not how wide it is.

The split depends on where the thresholds are placed, since the two curves are not related by a simple rescaling: measured between $20\%$ and $80\%$ the thermal share is $59\%$, and between $30\%$ and $70\%$ it is $61\%$. The qualitative statement, that the rounding is roughly half thermal, is stable across these choices; the value quoted above refers to the $10\%$--$90\%$ definition used throughout this work.

The athermal regime has a microscopic signature as well. We extract the Larkin correlation length in the pinned regime from the correlation of transverse displacements along the chain, $C(d)=\langle u_i u_{i+d}\rangle/\langle u_i^2\rangle$ with $u_i$ measured from the chain mean, taking $L_c$ as the separation at which $C$ falls to $1/e$.
This is the estimator used for Fig.~\ref{fig:micro}(a). Repeating it at fixed drive in the pinned regime ($F\le0.005$) while varying the temperature yields Fig.~\ref{fig:T0}(c). Up to $87$~K ($k_BT/V_0=0.039$) $L_c$ does not respond to temperature: five points from $T=0$ to $87$~K agree on $2.56\pm0.03$ bonds. From $145$~K upward it grows linearly, reaching $4.29\pm0.08$ at room temperature; the linear fit meets the plateau at $k_BT/V_0\approx0.03$ ($\approx60$~K), and the first point that lies clearly above the plateau is at $k_BT/V_0=0.065$ ($145$~K), so the onset of the thermal response lies between $k_BT/V_0\approx0.03$ and $0.065$.
The temperature scan of Sec.~\ref{sec:tscan} lies entirely above this threshold, at $k_BT/V_0=0.094$, $0.135$ and $0.229$, and so could not have seen the flat region.

The two observations are consistent with one another. The width that survives at $T=0$ and the correlation length that stops responding to temperature describe the same athermal regime, in which the chain conformation is set by the landscape and not by thermal fluctuations. The chain has an intrinsic correlation length of about two and a half bonds that disorder alone imposes; temperature lengthens it only once $k_BT$ reaches roughly a twenty-fifth of the well depth, $k_BT/V_0\approx0.039$.

It is worth being precise about what is sharp at $T=0$ and what is not. The onset of motion is sharp. At $F\leq 0.02$ the $N=200$ chain advances by less than $0.02$~nm over the entire run, a thousandth of a bond length, whereas at $F=0.035$ it covers $26$~nm, more than one bond: a factor of $10^3$ in displacement across one step of the drive grid. Below onset the chain is genuinely static, as it must be in the absence of thermal activation.

What is not sharp is the recovery of mobility above onset. Between the $10\%$ and $90\%$ marks the drive changes by $0.33$ decades within a single landscape, because segments that have cleared their pins are held back by segments that have not, and free flow is reached only when the last of them lets go. The athermal width measures that internal process rather than the threshold itself, which is why it is a property of one sample and why lengthening the chain does not reduce it. Thermal activation at $300$~K roughly doubles it.

\section{Discussion}

The one-third reduction of the depinning threshold, Eq.~\eqref{eq:onethird}, is a sizable and experimentally relevant effect. In a device that drives skyrmions through a disordered track, organizing the carriers into a chain rather than moving them as isolated textures lowers the threshold drive by a factor of three, with corresponding savings in dissipation. The size scan presented here shows that the effect is not a feature of a particular chain length but is instead a property of the collective elastic regime, valid already at $N=50$. The robustness of the ratio across $B=30$--$40$~mT, where the damping varies by more than a factor of two, further establishes that this is a structural, geometry-driven effect, not tied to the specific dissipation parameters of the texture. 

A quantitative account of the value $1/3$ follows from collective (Larkin) pinning~\cite{larkin1979}, provided the two reference forces are correctly identified. The drive is a force per unit length of the chain. Within one Larkin domain of $L_c$ bonds the pinning forces add incoherently, so the collective threshold per skyrmion is $F_c^{\rm chain}\sim f_{\rm rms}/\sqrt{L_c}$, with $f_{\rm rms}=V_0\sqrt{\eta_{\rm pin}}/\sigma$ the r.m.s.\ pinning force per skyrmion set by the Gaussian wells. Standard Larkin scaling with both thresholds referenced to this same r.m.s.\ force would therefore give $F_c^{\rm chain}/F_c^{(1)}=1/\sqrt{L_c}=0.5$ for the measured $L_c\approx 4$. 

The measured single-skyrmion threshold, however, is not set by $f_{\rm rms}$. The isolated skyrmion cannot avoid the strongest pins at this coverage and depins at the peak force of a single Gaussian pin, $F_c^{(1)}\simeq f_{\rm pin}=e^{-1/2}V_0/\sigma$ (the maximum gradient of the well), reproducing the measured $F_c^{(1)}$ of Table~\ref{tab:summary} to better than $8\%$ at the three fields and exceeding the r.m.s.\ force by $f_{\rm pin}/f_{\rm rms}=e^{-1/2}/\sqrt{\eta_{\rm pin}}\approx 1.8$ at our coverage $\eta_{\rm pin}\approx 0.118$. Using the appropriate reference for each object, r.m.s.\ for the chain and peak for the single skyrmion, the ratio is
\begin{equation}
\frac{F_c^{\rm chain}}{F_c^{(1)}}=\frac{f_{\rm rms}}{f_{\rm pin}\sqrt{L_c}}=\frac{\sqrt{\eta_{\rm pin}}}{e^{-1/2}\sqrt{L_c}},
\label{eq:ratioLarkin}
\end{equation}
where the collective advantage is $\sqrt{\rm coverage}$ over $\sqrt{\rm Larkin\ length}$: numerically $0.5/1.8\approx 0.28$ at all three fields, matching the measured $0.279$, $0.273$ and $0.273$. The independence from magnetic field follows because $\eta_{\rm pin}$ is fixed by construction and $L_c$ is measured to be field-independent.

The $T=0$ data test the two halves of this account separately, since $L_c$ and both thresholds are measured there as well. The collective half survives. With $L_c=2.56$ bonds at $T=0$ against $4.29$ at $300$~K, $f_{\rm rms}/\sqrt{L_c}$ predicts $F_c^{\rm chain}=0.045$ and $0.035$ respectively, against measured $0.0524\pm0.0003$ and $0.0354\pm0.0001$: ratios of $1.17$ and $1.02$. Both fall within the $\mathcal{O}(1/L_c)$ uncertainty discussed below, and the prediction tracks $L_c$ as it changes by a factor of $1.7$. The collective-pinning estimate is therefore verified at two temperatures rather than fitted at one.

The single-skyrmion reference does not survive. By construction $f_{\rm pin}$ contains no temperature, and it cannot describe a threshold that doubles when the temperature is set to zero: we measure $F_c^{(1)}=0.2588^{+0.0034}_{-0.0028}$ at $T=0$ against $0.1295\pm0.0030$ at $300$~K, that is $2.04$ and $1.02$ times $f_{\rm pin}$. At this coverage the wells overlap, and the deepest composite traps are stronger than any single Gaussian; at $T=0$ the isolated skyrmion needs nearly twice the single-pin peak force to escape one, while at $300$~K thermal activation brings the threshold back down to $f_{\rm pin}$. The agreement at room temperature is thus a coincidence and not a mechanism, and Eq.~\eqref{eq:ratioLarkin} should be read as accounting for the collective side of the ratio alone.

The ratio itself moves accordingly, from $0.273\pm0.006$ at $300$~K to $0.203\pm0.003$ at $T=0$. The collective advantage grows on cooling, because thermal activation lowers the isolated-skyrmion threshold by a factor of $2.0$ and the chain threshold by only $1.5$: a single skyrmion depends on thermal escape from one deep trap, whereas the chain already has a second route, redistributing force along its length, and gains less from being heated.

Three caveats justify a cautious reading of this agreement. The first is the one just described: the single-skyrmion reference is temperature-dependent in a way $f_{\rm pin}$ is not. Second, the collective-pinning account agrees with mean field only when $L_c\gg 1$. By this measure, $L_c\approx 4$ is not that large, so our estimate carries an $\mathcal{O}(1/L_c)\sim 25\%$ uncertainty. Third, there is an order-unity ambiguity in projecting the two-dimensional pinning force onto the drive direction. Using the drive-component variance $w^2/2$ in place of the full variance lowers Eq.~\eqref{eq:ratioLarkin} to $\approx 0.20$. Within these uncertainties we present Eq.~\eqref{eq:ratioLarkin} as accounting for the magnitude ($\approx 1/3$) and for the field-independence of the measured ratio, rather than as a precise prediction. The marginality of $L_c$ is itself consistent with the beyond-mean-field roughness $\zeta\approx 0.63$ (qKPZ) discussed below, which no mean-field estimate reproduces.

Two distinct sources of rounding contribute to the transition, and the $T=0$ comparison of Sec.~\ref{sec:T0} separates them; the size scan rules out a third. One might expect the longest-wavelength modes of the chain, whose wavelengths scale with $N$, to raise the depinning probability and so narrow the transition as the chain is lengthened. No such narrowing is observed: the width is flat from $N=50$ to $N=200$, and the coefficient of a $1/N$ term is consistent with zero. The rounding therefore cannot be attributed to finite-size modes at all, and must come from sources that do not scale with $N$: thermal activation over the pins~\cite{bustingorry2008,chauve2000}, and the athermal, single-realization mechanism of Sec.~\ref{sec:T0}. The thermal energy in our simulations, $k_BT/V_0=0.13$--$0.14$ at the three fields, is not negligible compared to the pin depth, and the resulting thermal activation rounds the transition even in the absence of finite-size effects. The width is not, however, independent of the applied field: as shown in Sec.~\ref{sec:fss}, the core of the transition widens by $26\%$ between $B=30$ and $40$~mT while $k_BT/V_0$ changes by only $5\%$. That field dependence is therefore not thermal in origin, and it is separate from the attribution made here, which concerns the rounding that survives at fixed field.
 We test this directly in Sec.~\ref{sec:tscan} by a temperature scan at fixed landscape (Fig.~\ref{fig:tscan}). The width grows monotonically with $k_BT/V_0$, with the rounding concentrated on the creep foot of the transition, which smears below the accessible drive at $T=510$~K while the bulk threshold $F_c^{(50)}$ shifts only mildly. The $T=0$ scan of Sec.~\ref{sec:T0} completes the accounting, showing that this thermal contribution is roughly half of the room-temperature width.

A consequence is that the classic critical depinning of an infinite elastic line~\cite{fisher1985,narayan1993}, with a sharp $F_c$ and a power-law $v\sim(F-F_c)^\beta$, is not realized here. Fits to $v=A(F-F_c)^\beta$ on the chain force the threshold to $F_c\to 0$ at every length $N$ and at every field $B$, suggesting that the depinning is a rounded crossover between two linear-mobility regimes (pinned with $\mu_{\rm pinned}\approx 0.02$--$0.04\,\mu_{\rm free}$ and free with $\mu_{\rm free}$), not a critical jump. The finite value of $w$ is the quantitative statement of this. In the flow regime, 
the measured roughness exponent is $\zeta\approx 0.61$--$0.71$ (with the fit window between $2$ and $8$ bond lengths), consistent with the quenched-KPZ value $\zeta_{\rm qKPZ}\simeq 0.63$ in $d=1$~\cite{tang1992,leschhorn1993} and clearly below the quenched-Edwards-Wilkinson depinning value $\zeta_{\rm qEW}\approx 1.25$~\cite{leschhorn1993}. The transient peak $\zeta\approx 0.88$--$0.89$ inside the crossover window belongs to neither class. It characterizes the strongly inhomogeneous, partially depinned state and we do not assign it a universality class. We note that the accessible scaling window is limited, since $L_c$ reaches up to $17$ bonds on a chain of $200$, and shifting the fit window by one bond at either end moves the quoted exponents by up to $\pm 0.1$.

The $N$-independence of the bulk threshold $F_c^{(50)}$, together with its weak field dependence ($F_c^{(50)}$ varies from $0.032$ at $B=30$ to $0.039$ at $B=40$, a change of $\approx 20\%$), has a simple physical interpretation. As with polymers, the relevant barrier for depinning is set by local properties of the skyrmion chain. These properties are governed by the bond energy, which determines $r_{eq}$; by the bending stiffness $\kappa_{\rm WLC}$; and by the dimensionless landscape parameters $V_0/D_e$, $\sigma/r_{eq}$ and the coverage. These are independent of the chain length. The mild dependence of $F_c^{(50)}$ on $B$ reflects the slight dependence of these local parameters on $B$. The universal one-third ratio reflects the fact that the same local parameters dictate both the chain and the isolated skyrmion thresholds proportionally.

Finally, our results do not explore the deep creep regime $F\ll F_c$. The classical creep law for a one-dimensional elastic line, $v\sim\exp[-(U_c/k_BT)(F_c/F)^{\mu}]$ with $\mu=(2\zeta_{\rm eq}+d-2)/(2-\zeta_{\rm eq})=1/4$ in $d=1$~\cite{chauve2000,kolton2009}, is derived under the assumption of harmonic elasticity. In a separate work~\cite{silva2026chains} we have shown that the transverse confinement of a skyrmion within the chain is \emph{quartic} rather than harmonic such that a small transverse displacement changes the bond length only at second order, so the harmonic radial stiffness of the bi-exponential bond generates a quartic transverse potential and a scale-dependent thermal exponent. Since the equilibrium roughness $\zeta_{\rm eq}$, and hence the creep exponent $\mu$, are derived from the harmonic elastic energy, the quartic transverse confinement of the skyrmion polymer raises the possibility of an anomalous creep law, with an exponent distinct from the universal $\mu=1/4$ of harmonic interfaces. This question is taken up in a companion work on the same calibration~\cite{silva2026creep}, which finds that thermal fluctuations generate an effective tension out of the quartic term, so that the equilibrium exponents of the disordered line, and with them the creep exponent, return to their harmonic values; the creep regime itself is shown there to be inaccessible to direct dynamics, for reasons of aging that lie outside the scope of the present study.

\section{Conclusion}

We have shown, by stochastic Thiele simulations with an atomistically derived bi-exponential interaction and periodic boundary conditions along the chain contour, that an elastic chain of magnetic skyrmions depins collectively at a drive that is close to one third of the single-skyrmion threshold, and that this ratio is (i)~a property of the bulk collective regime, valid already at $N=50$ and constant through $N=200$, and (ii)~robust against the applied magnetic field over $B=30$--$40$~mT. The depinning is a rounded crossover that proceeds, at $N=200$, through a sequence of internal events that begin with an internal deformation characterized by the Larkin length, the roughness exponent and the bond-survival fraction. This is followed by mobility recovery and Hall-angle recovery as the drive is increased. The width of the transition does not depend on $N$ over the range studied, so the rounding is not of finite-size origin. Repeating the size scan at $T=0$ separates the two contributions to it directly: the athermal limit retains $w(T=0)=0.334\pm0.004$ decades, so finite temperature ($k_BT/V_0\sim 0.1$) accounts for approximately half of the residual rounding and an athermal mechanism accounts for the other half. The same athermal regime is visible in the chain conformation: the Larkin correlation length in the pinned state is $2.56\pm0.03$ bonds and does not respond to temperature up to $k_BT/V_0=0.039$ ($87$~K), growing linearly only above a knee at $k_BT/V_0\approx0.03$--$0.06$. The onset of motion is nevertheless sharp: below it the chain advances by less than a thousandth of a bond length over an entire run, as it must in the absence of thermal activation. What the athermal width measures is not that onset but the range of drive over which mobility recovers afterwards, and that range belongs to a single disorder realization rather than to the ensemble.

The width does depend on the applied field, though not as a simple scale. Between $B=30$ and $40$~mT the core of the transition widens by $26\%$ while its creep tails shorten, so that widths read between different threshold pairs rank the three fields differently; the ratio $w_{10\text{--}90}/w_{20\text{--}80}$ falls monotonically from $1.87$ to $1.39$. Over the same range $k_BT/V_0$ changes by only $5\%$ while the damping falls by a factor of $2.4$, so this is a change in the shape of the transition driven by the dynamics rather than by the thermal scale.

Finally, the two temperatures test the collective-pinning account of the one-third ratio separately in its two halves. The collective side survives: $f_{\rm rms}/\sqrt{L_c}$ predicts the chain threshold to within $17\%$ at $T=0$ and $2\%$ at $300$~K while $L_c$ itself changes by a factor of $1.7$. The single-skyrmion reference does not: an isolated skyrmion needs $2.04$ times the peak force of one Gaussian well at $T=0$ against $1.02$ times at $300$~K, so the agreement of that reference at room temperature is a coincidence. The measured ratio moves from $0.273$ to $0.203$ on cooling, the collective advantage growing because thermal activation assists an isolated skyrmion more than it assists the chain. These results establish collective elasticity as a practical route to reducing the depinning threshold in devices using skyrmion chains, with a threshold reduction by a factor of three that is robust against the chain length and the operating field.

\section*{Data availability}

The data that support the findings of this article are available from the corresponding author upon reasonable request.

\begin{acknowledgments}
R. L. Silva and R. C. Silva acknowledge financial support from CNPq and FAPES (Grant No. TO1034/2025). R. L. Stamps acknowledges support from the Natural Sciences and Engineering Research Council of Canada (RGPIN 05011-18).
\end{acknowledgments}

\end{document}